\documentclass[acus]{JAC2000}

\usepackage{graphicx}

\begin{document}
\title{DESIGN AND PERFORMANCE SIMULATIONS OF THE BUNCH COMPRESSOR FOR THE
APS LEUTL FEL
\thanks{
Work supported by the U.S. Department of Energy, Office of Basic Energy
Sciences, under Contract No. W-31-109-ENG-38.}}

\author{M. Borland, ANL, Argonne, IL 60439, USA}

\maketitle

\begin{abstract} 

A magnetic bunch compressor was designed and is being commissioned to
provide higher peak current for the Advanced Photon Source's (APS)
Low-Energy Undulator Test Line (LEUTL) free-electron laser
(FEL) \cite{MiltonSASE}. Of great concern is limiting emittance growth due to
coherent synchrotron radiation (CSR).  Tolerances must also be
carefully evaluated to find stable operating conditions and
ensure that the system can meet operational goals.  Automated matching
and tolerance simulations allowed consideration of numerous
configurations, pinpointing those with reduced error
sensitivity.  Simulations indicate tolerable emittance growth up to
600 A peak current, for which the normalized emittance will increase
from 5 to about 6.8 $\mu$m.  The simulations also provide predictions
of emittance variation with chicane parameters, which we hope to
verify experimentally.
\end{abstract}

\section{INTRODUCTION}

A companion paper \cite{BorlandLINAC2K} reviews magnetic bunch
compression and shows a schematic of our system.  I assume the reader
is familiar with this paper.  The APS bunch compressor design is an
outgrowth of studies \cite{Emma} by P. Emma and V. Bharadwaj of
Stanford Linear Accelerator Center (SLAC).  They explored a number of
designs, including symmetric and asymmetric four-dipole chicanes.
Starting from this work, I investigated a large number of
configurations with various values of $R_{56}$, asymmetry, and final
current.  For each configuration, detailed longitudinal and transverse
matching was performed, followed by tracking with CSR and wakefields.
Then, sensitivity analysis was performed for all configurations,
followed by jitter simulations for the least sensitive configurations.

This work relied on {\tt elegant} \cite{elegant}, a 6-D code with a
fast simulation of CSR effects, plus longitudinal and transverse
wakefields.  {\tt elegant} also performs optimization of actual
tracking results, such as bunch length, energy spread, and emittance.

Simulation of the linac uses the {\tt RFCA} element, a matrix-based rf
cavity element with exact phase dependence.  Our linac has quadrupoles
around the accelerating structures. Hence, I split each 3-m section
into about 20 pieces, between which are inserted thin-lens,
2nd-order quadrupole elements.  A series of such elements is used for
each quadrupole.

A Green's function technique is used to model wakefields, using a
tabulation of the SLAC-structure wake functions provided by
P. Emma \cite{EmmaPC}. To reduce running time, one longitudinal wake
element is used per 3-m section, which is a good approximation for
relativistic particles.  For transverse wakes, I used one wake element
per rf cavity element (about 20 per section).

The CSR model used by {\tt elegant} is based on an
equation \cite{Saldin} for the energy change of an arbitrary line
charge distribution as a function of the position in the bunch and in
a bending magnet.  Details of this model will be presented by the
author at an upcoming conference.  Effects of changes in the
longitudinal distribution within a dipole are included.  CSR in drift
spaces is included by propagating the terminal CSR ``wake'' in each
bend through the drifts with the beam.

\section{MATCHING}

Longitudinal and transverse matching has the goal of providing
configurations for the 300-A and 600-A LEUTL operating
points \cite{BorlandLINAC2K}.  The starting point for the simulations
is macro particle data generated \cite{LewellenPC} with PARMELA, giving
the 6-D distribution after the photoinjector (PI).  See Figure 1 in
 \cite{BorlandLINAC2K}.

Longitudinal matching involves adjusting the phase and voltage of L2
(see  \cite{BorlandLINAC2K} for nomenclature) to obtain the desired
current and energy after the chicane.  Then, L4 and L5 are adjusted to
minimize the energy spread and obtain the desired final energy.
Longitudinal matching includes longitudinal wakefields, rf curvature,
and higher-order effects in the beam transport, by matching {\em
tracked} properties of the simulated beam.

\begin{figure}[htb]
\centering
\includegraphics*[width=80mm]{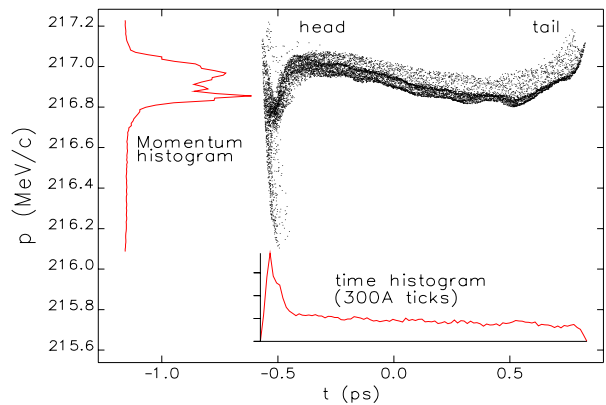}
\caption{Typical longitudinal phase space (300-A case)} 
\label{fig:longitDist}
\end{figure}

Figure \ref{fig:longitDist} shows the longitudinal phase space for the
300-A case with $R_{56}=-65$ mm, which exhibits a current spike of
nearly 1200 A.  The matching ignores this spike (which is shorter than
a slippage length for 530 nm) because of the way ``current'' is
defined \cite{BorlandLINAC2K}.

\begin{figure}[htb]
\centering
\includegraphics*[width=80mm]{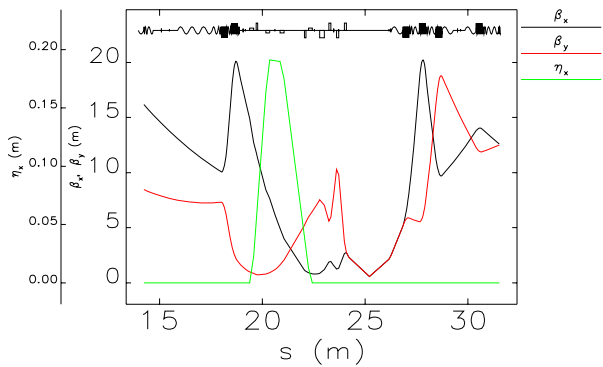}
\caption{Typical twiss parameters in the chicane region.} 
\label{fig:twiss}
\end{figure}

Following longitudinal matching, transverse matching is done for each
configuration.  Initial Twiss parameters are obtained from the rms
properties of the PARMELA beam.  Starting values for the quadrupoles
were obtained from matching ``by hand'' for one configuration.
Four sequential {\tt elegant} runs work the beta functions down the
linac.  The most important constraints maintain small beta functions
in the linac (for transverse wakefield control), small horizontal beta
in dipole B4 (to reduce CSR effects), and matching for the
emittance measurement sections.  Figure \ref{fig:twiss} shows sample
Twiss parameters in the chicane region.

The matching is highly automated, so that only the desired beam
current and energy needs to be specified.  Evaluation of tolerances
and randomized simulations are also automated, being set up by scripts
from the corresponding matching runs.  Transfer of data between
simulation stages is handled using SDDS files and
scripts \cite{BorlandICAP98}, reducing errors and increasing the number
of configurations that can be examined.  For example, a script is used
to scan all configurations and give power supply specifications.  A
distributed queue utilizing 50 workstations is used to run the
simulations.

\begin{figure}[htb]
\centering
\includegraphics*[width=80mm]{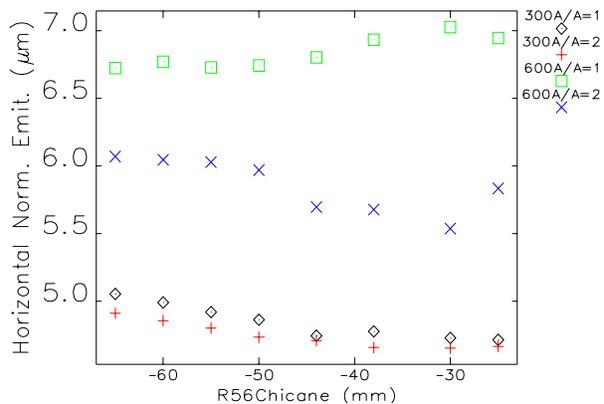}
\caption{Horizontal normalized emittance vs. $R_{56}$.} 
\label{fig:emitTrends}
\end{figure}

Figure \ref{fig:emitTrends} shows emittance vs. $R_{56}$ for the
symmetric (A=1) and asymmetric (A=2) cases at 300 A and 600 A.  For
300 A, the symmetric and asymmetric cases are very similar.  For 600
A, the difference is 10\% or more, which should be measurable.

One surprise in Figure \ref{fig:emitTrends} is that the emittance does
not uniformly increase as $|R_{56}|$ increases, even though {\tt
elegant} shows the expected monotonic increase (due to CSR) vs bending
angle for a single dipole with a constant input beam distribution.
This is apparently due to variation in the compressed bunch
distribution between cases with the same ``current'' but different
$R_{56}$.  For smaller $|R_{56}|$, there are higher current spikes at
the head of the bunch, leading to a larger and more rapidly changing
CSR wake, which in turn leads to larger emittance growth.  The effect
is even more pronounced in the 1200-A cases (not shown).  Insertion of
the scraper between B2 and B3 to remove the low-energy part of the
beam can reduce the height and width of the current spike, resulting
in lower emittance.  Unfortunately, this also reduces the current in
the rest of the bunch considerably.

Earlier simulations showed that emittance trends can be changed
significantly by inconsistent values of the horizontal beta function at
the exit of B4.  All of these subtleties will make for difficult
interpretation of experiments in which $R_{56}$ is varied.  However,
because compression to different currents for fixed $R_{56}$ involves
only adjustment of the rf phases and voltages, comparision of the
emittance growth for different amounts of compression should be more
straightforward.

\section{TOLERANCE DETERMINATION}

Tolerances are driven by the FEL gain length, trajectory, and
wavelength stability requirements \cite{MiltonPC}.  The 10\% rms gain
length variation limit is easy to use in {\tt elegant} as it computes
FEL performance directly using Xie's parameterization \cite{MXie}.
Beam trajectory limits ($\sim$50 $\mu$m, $\sim$50$\mu$r) are included
separately as they are not incorporated into Xie's formula.  The 1-nm
rms wavelength variation limit is a challenging goal at 530 nm as it
puts a 0.1\% limit on energy variation.

The analysis begins by running single-parameter ``sweeps'' to assess
the effect on the constrained quantities (gain length, trajectory, and
wavelength) of accelerator parameters (e.g., rf phase).  Sweeps
included rf phase and voltage; photoinjector timing, charge, and
energy; and chicane dipole strength.  From these sweeps, a script
determines the limit on each parameter change due to the various
specifications, showing that configurations with the largest $R_{56}$
are least sensitive to difficult-to-control timing and phase errors.
These configurations experience the most emittance degradation from
CSR, but tend to yield the shortest gain length as they have the
smallest energy spread (L2 being closer to crest).

\begin{table}[h] 
    \caption{Selected sweep limits for $R_{56}=-65$ mm}
    \label{tab:sweepLimits}
    \begin{center}
        \begin{tabular}{|l|c|c|}\hline
        quantity        &  300-A limit & 600-A limit \\ \hline
        L2 phase        &  0.17$^\circ$   & 0.49$^\circ$ \\ \hline 
        L4/L5 phase     &  0.77$^\circ$   & 1.45$^\circ$ \\ \hline
        L2 voltage      &  0.11\%      & 0.31\% \\ \hline
        L4/L5 voltage   &  0.52\%      & 1.4\% \\ \hline
        PI timing       &  0.29 ps     & 0.88 ps  \\ \hline
        PI energy       &  0.26\%      & 1.1\%  \\ \hline
        PI charge       &  12\%        & $>$20\%  \\ \hline
        \end{tabular}
    \end{center}                                    
\end{table}

The limits, shown in Table \ref{tab:sweepLimits}, are larger for the
600-A case because the 1-nm wavelength constraint is easier at 120
nm than 530 nm.  Nine parameters are limited primarily by the
wavelength constraint and four others by horizontal trajectory
constraints. Hence, to determine the rms tolerance, one simply divides
each sweep limit by $\sqrt{N}$, $N$ being the number of parameters
limited by a particular constraint.  For the horizontal trajectory,
$N$ was doubled to eight to allocate half the budget to nonswept
parameters (e.g., corrector magnets).  Some of these phase and timing
tolerances are beyond the state of the art.

\section{RANDOMIZED SIMULATIONS}

Randomized simulations were used to confirm the tolerances and examine
errors not covered by the sweeps (e.g., corrector jitter, quadrupole
jitter, and alignment).  These were done for the most stable
configurations (i.e., $R_{56}=-65$ mm).  Because some tolerances are
beyond the state of the art, I used randomized simulations to
determine the impact of ``relaxed'' tolerances, assuming these rms
levels \cite{TravishPC}: 1$^\circ$ rf phase jitter, 0.1\% rf voltage
jitter, 1 ps timing jitter, 5\% charge jitter, and 2\% PI energy
jitter.

Tables \ref{tab:ranSweep} and \ref{tab:ranRelaxed} show the results,
respectively, for the sweep-derived tolerance levels and the relaxed
levels.  The sweep-derived tolerance levels result in meeting the
specifications for the FEL, while the relaxed levels, not
surprisingly, do not.  One surprise in the relaxed case is the large
jitter in the vertical plane.  This results from uncorrected nonlinear
dispersion in a vertical dogleg between the linac and the LEUTL, a
problem which can be readily remediated using two sextupoles 
\cite{EmmaPC}.

\begin{table}[h] 
    \caption{Results of 300 randomized simulations with sweep-determined tolerance levels for $R_{56}=-65$ mm}
    \label{tab:ranSweep}
    \begin{center}
        \begin{tabular}{|l||c|c|c|c|}\hline
               & \multicolumn{2}{|c|}{300 A} & \multicolumn{2}{|c|}{600 A} \\ \hline
          quantity                 & rms        & \%     & rms    & \%  \\ 
                                   & jitter     & inside & jitter & inside \\ \hline
$\langle x\rangle$ $(\mu m)$       &  71        & 83     & 57     & 91           \\ \hline
$\langle x\prime\rangle$ $(\mu r)$ &  29        & 93     & 24     & 96           \\ \hline
      $\langle y\rangle$ $(\mu m)$ &  13        & 100    & 11     & 100          \\ \hline
$\langle y\prime\rangle$ $(\mu r)$ &  19        & 98     & 17     & 99           \\ \hline
                  $L_{gain}$ $(m)$ &  0.01      & 99     & 0.016  & 100          \\ \hline
                  $\lambda$ $(nm)$ &  0.83      & 72     & 0.29   & 100          \\ \hline
        \end{tabular}
    \end{center}                                    
\end{table}

\begin{table}[h] 
    \caption{Results of 300 randomized simulations with relaxed tolerance levels for $R_{56}=-65$ mm}
    \label{tab:ranRelaxed}
    \begin{center}
        \begin{tabular}{|l||c|c|c|c|}\hline
               & \multicolumn{2}{|c|}{300 A} & \multicolumn{2}{|c|}{600 A} \\ \hline
          quantity                 & rms        & \%     & rms    & \%  \\ 
                                   & jitter     & inside & jitter & inside \\ \hline
$\langle x\rangle$ $(\mu m)$       &  89        & 72     &  89    & 81           \\ \hline
$\langle x\prime\rangle$ $(\mu r)$ &  59        & 64     &  68    & 58           \\ \hline
      $\langle y\rangle$ $(\mu m)$ &  63        & 88     & 127    & 79           \\ \hline
$\langle y\prime\rangle$ $(\mu r)$ &  138       & 62     & 245    & 39           \\ \hline
                  $L_{gain}$ $(m)$ &  0.048     & 68     &   3    & 1.3          \\ \hline
                  $\lambda$ $(nm)$ &  9.6       &  9     & 2.8    & 27           \\ \hline
        \end{tabular}
    \end{center}                                    
\end{table}

\section{ACKNOWLEDGEMENTS}

The technical note \cite{Emma} by P. Emma and V. Bharadwaj provided a
valuable starting point.  I acknowledge helpful discussions and
assistance from H. Friedsam, E. Lessner, J. Lewellen, S. Milton, and
G. Travish.  J. Lewellen provided the PI beam distribution data.

\end{document}